\documentclass{vgtc}                          % final (conference style)
\graphicspath{{figures/}{pictures/}{images/}{./}} % where to search for the images

\usepackage{times}                     % we use Times as the main font
\usepackage{tabu}                      % only used for the table example
\usepackage{booktabs}                  % only used for the table example
\usepackage{lipsum}                    % used to generate placeholder text
\usepackage{mwe}                       % used to generate placeholder figures

\usepackage{mathptmx}                  % use matching math font

\usepackage{soul}
\usepackage{tcolorbox}
\usepackage{subcaption}

\onlineid{0}

\vgtccategory{Research}

\vgtcinsertpkg

\title{``Trust Junk'' Leads to Unjustified Support \\for Highly Discriminatory Predictive Models}

\author{
Michael Correll\thanks{e-mail: m.correll@northeastern.edu} \\%
\scriptsize Northeastern University
\and Lucy Havens\thanks{e-mail: l.havens@northeastern.edu} \\%
\scriptsize Northeastern University
\and Mahsan Nourani\thanks{e-mail: m.nourani@northeastern.edu} \\%
\scriptsize Northeastern University
}

\teaser{
  \centering
  \includegraphics[width=\linewidth]{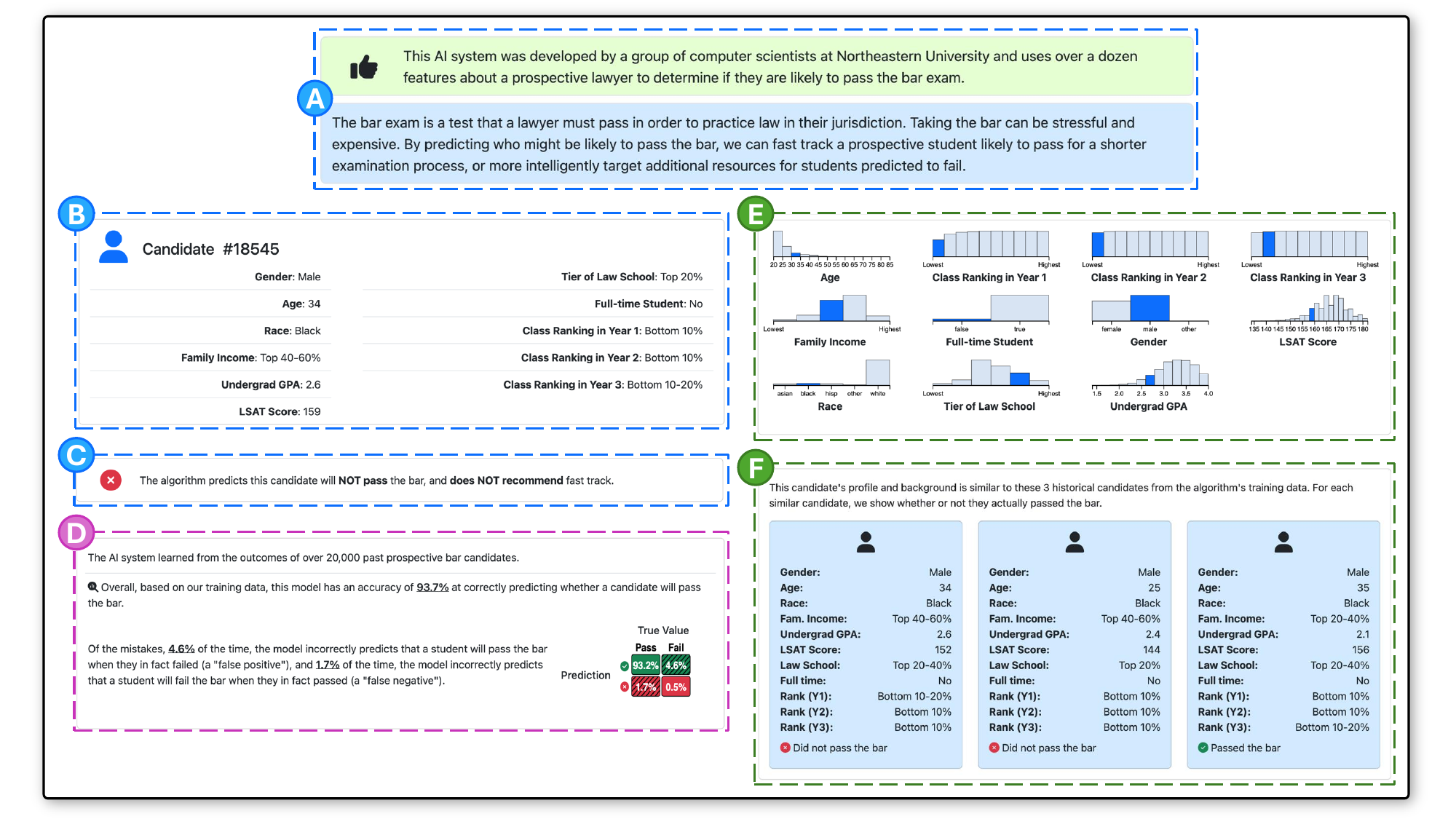}
  \caption{An annotated version of a stimulus from our experiment. A decision (\boxC) made by an automated system on whether or not a potential candidate (\boxB) is likely to pass the bar exam, along with explanatory information that serves as \textit{trust junk}~\cite{wall2024trust} (\boxA\boxD\boxE\boxF). The model is highly discriminatory (it predicts failure for candidates who identify as Black men, and predicts all other candidates will pass). Yet, providing increasing amounts of (largely fairness-irrelevant) explanatory information makes participants agree with the model more, and be more likely to rate the model as fair and unbiased. We empirically demonstrate that trust junk can mislead XAI viewers even when the explanatory information it provides is not, strictly speaking, incorrect.
  }
  \label{fig:teaser}
}

\abstract{
The persuasive power of data visualizations can go awry: for instance, in an explainable AI (XAI) context, visualizations can produce \textit{over-trust} of predictive models. In this paper, we use a crowdsourced study to show that providing accurate (but superfluous or irrelevant) data in a model explanation can, in fact, result in unjustified trust and other positive beliefs about a model, even when the model is patently discriminatory and unfair. Our results suggest that XAI designers and developers need to consider the implicit or explicit rhetorics of their work, and beware of the potential of visualizations to imbue models with unearned trust.
\newline
\newline
\textbf{Supplemental Material} is available at \href{https://osf.io/wufqz/}{https://osf.io/wufqz/}.
}

\keywords{XAI, Data Rhetoric, Information Visualization}

\begin{document}

\newcommand{\mahsan}[1]{{\color{magenta}{MN: #1}}}
\newcommand{\mc}[1]{{\color{purple}{MC: #1}}}
\newcommand{\lh}[1]{{\color{blue}{LH: #1}}}
\newcommand{\TODO}[1]{\textcolor{red}{TODO: #1}}
\newcommand{\edit}[1]{\textcolor{red}{#1}} %comment this and uncomment the next line before submission.

\newcommand{\bullshit}{bullshit}

\definecolor{circleBackground}{RGB}{38,169,253}
\definecolor{circleLines}{RGB}{13,110,253}

\newtcbox{\circleBlue}{on line,
  colback=circleBackground,
  coltext={white},
  colframe=circleLines,
  boxrule=0.5pt,
  arc=5pt,
  boxsep=0pt,
  left=2.5pt,
  right=2.5pt,
  top=2pt,
  bottom=2pt,
% halign=center,valign=center,
% square,circular arc
}

\definecolor{circlePink-Bg}{RGB}{216,110,204}
\definecolor{circlePink-Lns}{RGB}{200,51,186}

\newtcbox{\circlePink}{on line,
  colback=circlePink-Bg,
  coltext={white},
  colframe=circlePink-Lns,
  boxrule=0.5pt,
  arc=5pt,
  boxsep=0pt,
  left=2.5pt,
  right=2.5pt,
  top=2pt,
  bottom=2pt,
% halign=center,valign=center,
% square,circular arc
}

\definecolor{circleGreen-Bg}{RGB}{78,167,46}
\definecolor{circleGreen-Lns}{RGB}{26,107,37}

\newtcbox{\circleGreen}{on line,
  colback=circleGreen-Bg,
  coltext={white},
  colframe=circleGreen-Lns,
  boxrule=0.5pt,
  arc=5pt,
  boxsep=0pt,
  left=2.5pt,
  right=2.5pt,
  top=2pt,
  bottom=2pt,
% halign=center,valign=center,
% square,circular arc
}

\newcommand{\condBase}{\textcolor{circleBackground}{\textbf{Baseline}}}
\newcommand{\condHalf}{\textcolor{circlePink-Bg}{\textbf{Model}}}
\newcommand{\condAll}{\textcolor{circleGreen-Bg}{\textbf{Everything}}}

\newcommand{\boxA}{\circleBlue{\textbf{A}}}
\newcommand{\boxB}{\circleBlue{\textbf{B}}}
\newcommand{\boxC}{\circleBlue{\textbf{C}}}
\newcommand{\boxD}{\circlePink{\textbf{D}}}
\newcommand{\boxE}{\circleGreen{\textbf{E}}}
\newcommand{\boxF}{\circleGreen{\textbf{F}}}


\firstsection{Introduction}

\maketitle
As articulated by Kennedy et al.~\cite{kennedy2016work}, data and their presentations can produce the (false) impression of ``objectivity'' as well as ``transparency, scientific-ness and facticity'' through forms of implicit rhetorical work. Likewise, Peck et al.~\cite{peck2019data} find that mass audiences often have intrinsic trust in data visualizations even from sources they think of as biased in other contexts. Therefore, Drucker~\cite{drucker2012humanistic} argues we must beware of the ``persuasive and seductive rhetorical force of visualization.'' The unjustified authority of data, and data visualizations, is highly pertinent to the design of XAI. A particular danger in XAI is if machine learning (ML) system explanations that appear authoritative and complete ``soothe''~\cite{weller2017challenges} viewers while failing to provide information needed to meaningfully assess the system's accuracy, fairness, or performance. Wall et al.~\cite{wall2024trust} expand on this threat through their concept of \textbf{trust junk}, where an XAI visualization ``that has no meaningful connection with the underlying model or data is employed to enhance trust.'' %producing a ``disconnect between information intended to improve a user’s trust in the model and the model itself.'' 
In other words, explanations with trust junk may include visualizations that persuade by being beautiful, complicated, or detailed without being \textit{useful}. Concerningly, trust junk may therefore \textbf{fairwash}~\cite{aivodji_fairwashing_2019} models--- in other words, produce unearned and unwarranted assumptions of trust or fairness in models that are, in reality, biased and inaccurate.

In this paper, we report a crowdsourced study exploring the impact of trust junk in ML model explanations. We find that, even for an intentionally unfair model, increasing the amount of seemingly useful (but in actuality irrelevant) data in an explanation increases perceived user trust and agreement with the model, while also \textit{reducing} perceptions of bias or unfairness. These results suggest that XAI designers must take accountability not only for the accuracy of the data used in explanations, but also for the persuasive force of the explanations. Even without an intent to deceive, XAI techniques can fall into ``explainability pitfalls'' when considered in their context of use and interpretation by their intended audiences~\cite{ehsan2024explainability}.
%These results suggest that designers working in XAI should consider not merely the truth or falsity of the data they present in their explanations, but also the persuasive work done by these presentations of data.

\section{Background}
The goals and techniques associated with XAI are vast (as demonstrated in numerous surveys~\cite{adadi2018peeking,Al-Ansari_2024,mohseni2021multidisciplinary,Hamida_2024,Kalasampath_2025,hoffman2018metrics,Schwalbe_Finzel_2024}).
However, determining how to create effectively human-centered explanations remains an open area of study.
Researchers have investigated how to choose the appropriate level of detail~\cite{kulesza_too_2013, liao2020questioning,poursabzi2021manipulating}, type of explanation~\cite{laato2022explain}, and whether to show explanations at all~\cite{nourani2019effects}.

While there is growing work on intentionally \textit{adversarial} XAI~\cite{lakkaraju2020fool,dimanov2020you,pruthi2020learning,slack_fooling_2020} (e.g., gaming fairness scores or explanatory metrics to hide unfavorable model information) and ``dark patterns''~\cite{ehsan2024explainability}, we focus on a wider space of sociotechnical failures and the extent that explanations, even accurate ones, can work rhetorically to ``fairwash''~\cite{aivodji_fairwashing_2019} biased models by imparting them with (perhaps unearned) authority.
XAI techniques can exacerbate automation bias~\cite{bussone2015role,jacobs2021machine, bussone2015role, nourani2021anchoring}---a cognitive bias characterized by an overreliance on automated systems. Merely referring to an AI system as a ``Statistical Model'' instead of an ``Artificial Intelligence'' model can impact the perceived complexity and competency of the system~\cite{langer_look_2022}. Other effects are more subtle. Cabitza et al.~\cite{cabitza2024explanations} observe the ``XAI halo effect,'' where a high- or low-quality explanation may induce users to think of a model as being of similar quality as its explanation, even with insufficient evidence to support this judgment. Of particular concern to our work are ``placebic''~\cite{eiband2019impact} and ``empty''~\cite{weller2017challenges} explanations, which contain no useful information but can persuade by giving the appearance of being informative. Our study looks at an intentionally extreme case, measuring how increasing the amount of information in an explanation can persuade a viewer to trust a model more than they should.% while hiding its flaws.

\section{Motivating Scenario \& Techniques}
\label{sec:techniques}

Our user study design is based on controversial uses of AI systems to assign students' grades on the International Baccalaureate (IB) and General Certificate of Education (GCE) Advanced (A) Level exams~\cite{Ehsan_2022,ONeil_2016}.
We use the \textit{Law School Admissions Bar Passage} dataset~\cite{wightman1998lsac}, which contains demographic, academic, and bar exam result information for 20,000 candidates who took the bar exam from 1991--1997, to generate an intentionally biased model. Our model predicts failure for Black male candidates and passes for all other candidates. Yet, due to class imbalances in exam outcomes across race and gender, the model has 93.7\% accuracy. %, and low false positives and false negatives (4.6\% and 0.5\% of the training data, respectively).  

% \begin{enumerate}
Wall et al.~\cite{wall2024trust} claim that certain strategies such as communicating provenance information, providing transparency, or increasing the amount of data views in an XAI visualization can foster trust. This trust can be engendered regardless of the actual capabilities of the model: performing \textit{too much} trust-building is therefore akin to turning an ``evil knob'' too far, raising the risk of ``fair-washing''~\cite{aivodji_fairwashing_2019} a biased or inaccurate model.
Based on techniques described, but not tested, in Wall et al., as well as adversarial XAI work, we created ``junk'' explanations for our user study.  The explanations are themed around four techniques that we find to be misleading, even as they present data that is correct.% and accurate.

\noindent
\textbf{Burying in Details}: We provided an excessive amount of information for our model's binary classification task. Our inclusion of large amounts of complex-looking but ultimately irrelevant details was intended to overwhelm user study participants rather than provide genuinely useful data. This technique, which metaphorically numbs the user into accepting their own inability to fully understand the data, is referred to by Correll~\cite{correll_towards_2021} as a ``novocaine chart.''

%acting as a ``spectacle''~\cite{spectacular,gregg2015inside} rather than genuinely useful data.

\noindent
\textbf{Appeals to Authority}: We provided superficially impressive information, such as the name of the prestigious university where computer scientists developed the model (\autoref{fig:teaser}, \boxA), dataset size, and the model's accuracy score. This strategy is especially salient when the audience lacks a baseline: in our case, the accuracy of our model, at  $93.7\%$, is \textit{lower} than the accuracy of a model that would have predicted \textit{all} candidates passed regardless of their background, which would be $94.8\%$. Our choice to describe the prediction model as an ``AI system'' is also meant to encourage the reader to ascribe undue complexity or accuracy~\cite{langer_look_2022} to our model that is, at heart, a glorified if statement. 

\noindent
\textbf{Cherry Picking}: Many metrics reveal the unfair reliance of our model on gender and race, such as a $\chi^2$ test and feature importance scores.  We intentionally omitted these, instead providing metrics that represent our model's performance favorably (\autoref{fig:teaser}, \boxD).
We also provided a cohort-based explanation by showing the three most similar candidates to the main candidate based on Euclidean distance and displayed their \textit{ground truth} exam outcomes from the training data, rather than the \textit{model's predictions} (\autoref{fig:teaser}, \boxF). This allowed us to better hide that all Black men would be predicted to fail, even though only a fraction did so in reality.
%We also chose a small number of examples, making a reliable per-group contrastive analysis of systematic differences difficult.

\noindent
\textbf{Encouraging Folk Algorithms}: In the absence of knowledge of algorithmic internals, people often develop ``folk algorithms''--- simplified and often incorrect understandings about how an algorithm operates~\cite{ytre2021folk}. To encourage the creation of inaccurate folk algorithms, we provided information about \textit{all} features in the training data (\autoref{fig:teaser}, \boxB, \boxE, and \boxF), implying that the model used all of them in its decision-making process, when it only used 2 (i.e., race and gender). Likewise, by showing that a candidate was particularly high or low in certain attributes, our explanations encouraged participants to create inaccurate causal stories (which are particularly pernicious in visualizations of relationships) about why the model made a particular prediction~\cite{xiong2019illusion}.

\section{User Study}\label{sec:evaluation}
We conducted a between-subjects crowdsourced experiment on ``trust junk'' in XAI explanations, investigating whether this ``junk'' could persuade users that a biased model was fair and useful. Our main manipulation was to increase the number of explanatory components, where each component was technically \textit{correct} but unhelpful for assessing model efficacy or fairness (see \autoref{sec:techniques}). This manipulation is what Wall et al.~\cite{wall2024trust} refer to as a ``knob'' that designers can manipulate to impact trust in a model. While our stimuli are inspired by real XAI visualizations (our confusion matrix is based on work by Gomez et al.~\cite{gomez2021advice} and our cohort-based explanations are based on the ``C-NN visual explanations'' of Szymanski et al.~\cite{szymanski_visual_2021}), the resulting XAI dashboards are our own design, embodying a variety of explanatory techniques from prior work. 

Each of our explanations had one of three levels of trust junk corresponding to one of three study conditions.
% \noindent
Participants in the \textbf{\condBase{}} condition saw an explanation with basic model provenance information, a bar exam candidate's profile, and the model's decision for that candidate (\boxA, \boxB and \boxC).
% \noindent
In the \textbf{\condHalf{}} condition, participants saw an explanation with model accuracy statistics (\boxD) in addition to all of the information in the \textbf{\condBase{}} condition.
% \noindent
In \textbf{\condAll{}}, participants saw an explanation that, in addition to all information in the prior two conditions, also includes histograms of the candidate's feature scores (\boxE) and short profiles of similar candidates in the training data (\boxF).

The study was approved by our institutional review board and conducted on Prolific via Qualtrics. Participants' were compensated $\$15/$hour. 
Participants were randomly assigned to one of the three conditions. Their main task was to review eight profiles of bar exam candidates, presented in random order, and judge whether each candidate would pass the bar. We used this task to measure participants' agreement with the model as a reliance metric~\cite{nourani2020don,poursabzi2021manipulating}).
Then, participants responded to post-study questionnaires, adapted from previous work~\cite{goyal_impact_2024,hoffman2018metrics}, so we could assess their perceptions of and trust in the explanations and model predictions. After removing data from participants who failed our attention check, our final sample included 28, 27, and 28 participants in the \condBase{}, \condHalf{}, and \condAll{} conditions, respectively.

We hypothesized that increasing amounts of seemingly detailed (but ultimately distracting or at least incomplete) information will ``fairwash''~\cite{aivodji_fairwashing_2019} our model by fostering unearned and unwarranted trust in it. 
%and that this effect will be exacerbated as the amount of trust junk increases (akin to turning the ``evil knob'' too far~\cite{wall2024trust}). We particularly focus on a specific set of trust junk claims: that communicating provenance information, (seeming) transparency, and more data views in an XAI visualization fosters trust \textit{orthogonally} to the actual performance and fairness of the model. While Wall et al.~\cite{wall2024trust} claim other factors such as social influences and interactivity can also foster unearned trust, for experimental parsimony we focus on those factors amenable to modifying a static display within the bounds of a single crowd-sourced experiment.
%lead to a ``fairwashing'' outcome}: that is, 
More specifically, for participants in the \condAll{} and \condHalf{} conditions, relative to those in our \condBase{} condition, we hypothesized there will be (1) greater agreement with the model, (2) greater perception that the model is fair and trustworthy, and (3) greater perception that the model and explanation are useful.
%To these these hypotheses, we opted for mixed-method analyses. For quantitative measures, we used independent one-way ANOVAs to test if there are any significant differences across the three conditions, and used Tukey's HSD post-hoc test when necessary, for pairwise comparison of those conditions. 
% \subsection{Analysis Methods}
% Except where noted, we used independent one-way ANOVAs of the impact of the condition assignment on the variable of interest, with one data point per participant. When necessary, we calculated aggregate measures from per-trial measures (e.g., model agreement is measured as the number of times the participant-predicted bar passage outcome matched the model's prediction across all eight presented candidates). In cases of significant effect, we conducted a post-hoc Tukey's test of Honest Significant Difference (HSD) for pairwise comparison and report conditions with significant differences. For graphical results of continuous measures (such as agreement but also scale results made up of multiple Likert-scale items), we calculate per-condition means with error bars representing $+/-$ one standard deviation. For categorical measures (such as individual rating items, where aggregation into a single value might be misleading or incomplete~\cite{south2022effective}), we augment statistical tests with stacked bar charts of response frequency.
To understand participants' rationale for their responses throughout our user study, and gain insight into their mental models %, and in particular the resulting ``folk algorithms'' they built up of the model's internals, 
and ``folk algorithms'' of our model, we analyzed their free text responses to questions around model performance and fairness using qualitative coding. We were interested in whether participants noticed that the model was discriminatory with respect to race or gender, or that the model explanations were incomplete.

Additional study details including the full survey instrument, stimuli, analyses, participant demographics, and qualitative coding procedures are available in our supplement at \href{https://osf.io/wufqz/}{https://osf.io/wufqz/}.
%This includes the full survey instrument, stimuli, and participant demographics, as well as our R Markdown code for statistical tests, figure generation, and the full qualitative analysis procedure.

% Each of these concerns was coded with one three values: ``yes'' (if the participant explicitly addressed one of these concerns), ``no'' (if they did not), or ``unsure'' (if the participant might have been addressing one of these concerns, but there wasn't enough information to make a reliable judgment).
% Two coders independently coded each participant, including those who participated but failed our attention checks.
% After this process, the coders met to discuss mismatches (10/81 codes for noting bias issues around race or gender, 19/81 codes for noting missing or incomplete information in the explanation) and arrive at a consensus code.

\section{Results}
Here, we provide an overview of our main quantitative and qualitative results, broken down by measure.

%\begin{figure}
%    \centering
%    \includegraphics[width=0.75\columnwidth]{figures/genderDemo.pdf}
%    \caption{The distributions of participants based on their self-reported gender across the three study conditions.}
%    \label{fig:gender-distribution}
%\end{figure}

\begin{figure*}
    \centering
    \begin{subfigure}{0.2\textwidth}
        \includegraphics[width=\linewidth]{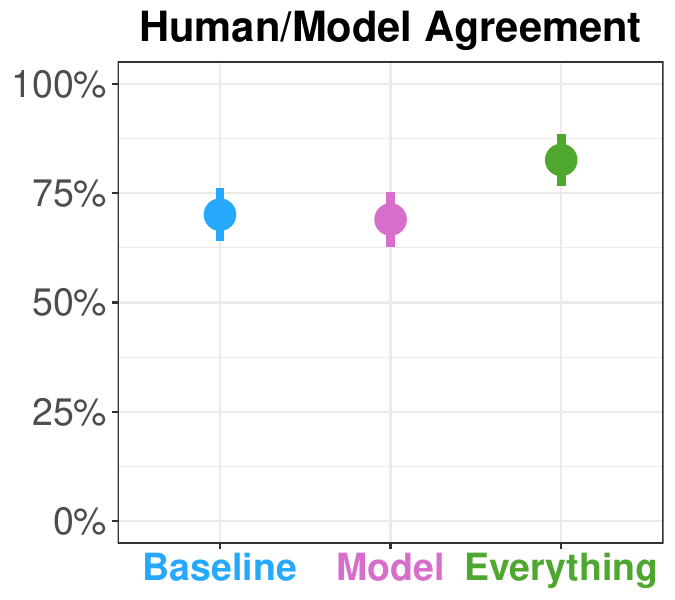}
        \phantomsubcaption
        \label{fig:agree}
    \end{subfigure}
    \begin{subfigure}{0.2\textwidth}
        \includegraphics[width=\linewidth]{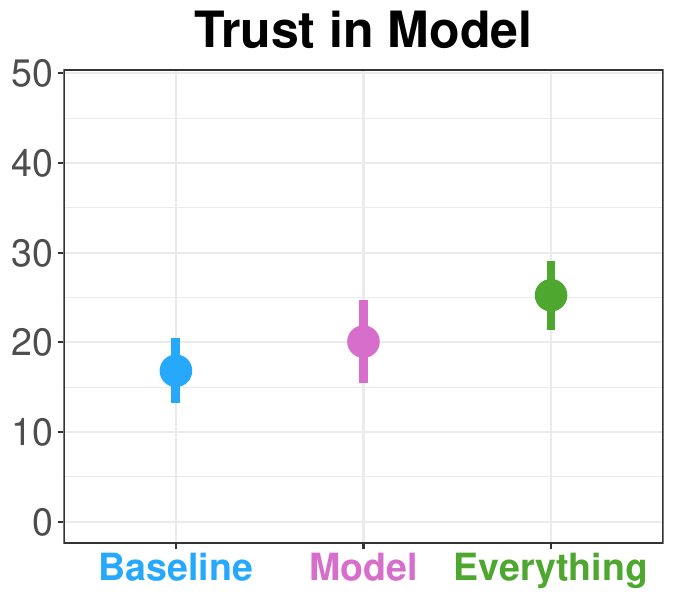}
        \phantomsubcaption
        \label{fig:qualitya}
    \end{subfigure}
    \begin{subfigure}{0.2\textwidth}
        \includegraphics[width=\linewidth]{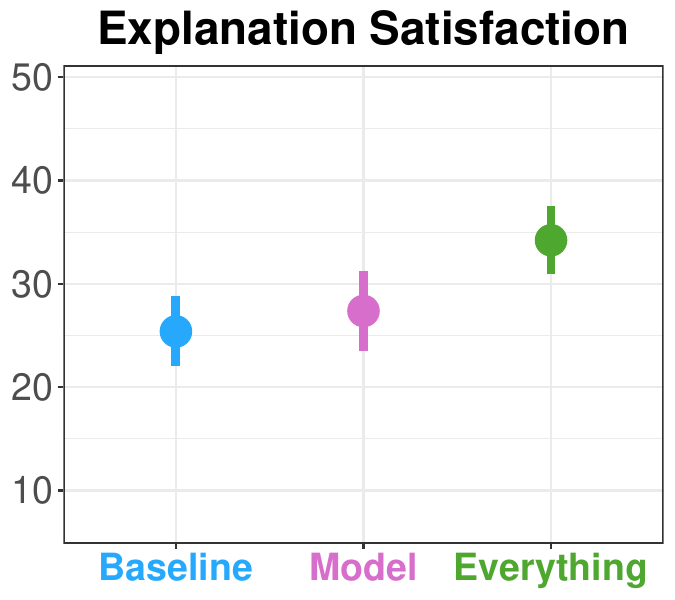}
        \phantomsubcaption
        \label{fig:quality}
    \end{subfigure}
    \vspace{-1em}
    \caption{Aggregate scores from our scales of agreement with (left) and perceived trust in (middle) the \textit{AI model}, as well as satisfaction \textit{with} the \textit{explanatory information} (right), based on scales used in Hoffman et al.~\cite{hoffman2018metrics}. Participants in the \condAll{} condition, who were provided with information that was ultimately insensitive to fairness assessments, had significantly higher ratings even though the underlying model was identical (and patently unfair) in all conditions. Error bars are 95\% t-confidence intervals of the mean.}
    \label{fig:quant}
\end{figure*}

% \subsubsection{Quantitative Analyses}
\noindent
\textbf{Model Agreement and Task Performance}: Participants correctly predicted the ground truth label $5.3/8$ times ($66.6$\% of the time) and  
%However, in keeping with the automation bias observed in prior work~\cite{goddard2012automation}, the average participant 
agreed with the model $5.9/8$ times ($73.9$\% of the time). Condition had a significant impact on agreement ($F(2,80)=6.4$, $p=0.0025$): %, with a post-hoc test finding that 
participants in the \condAll{} condition followed the model significantly more often ($M = 82.6$\%) than the other conditions ($M = 69.5$\%) (see \autoref{fig:agree}). %When participants' predictions differed from the ground truth, this was mostly ($M=80.0$\%) due to them following the model. 
Condition also had a significant impact on rate of over-reliance ($F(2,80)=5.1$, $p=0.008$): %with a post-hoc test finding that 
participants in the \condAll{} condition were more likely to erroneously follow the model ($M = 89.3$\% of errors) compared to the other conditions ($M = 75.3$\% of errors).

\noindent
\textbf{Model Trust}: 
% We summed the eight component Likert scale items, except for one negatively coded item, ``I am wary of the algorithm,'' which we subtracted, into a single value ranging from 0-48. 
Condition had a significant impact on the Likert scale trust rating ($F(2,80)=4.7$, $p=0.0.012$). A post-hoc test found that participants in the \condAll{} condition rated higher trust in the model ($M = 25.3$) than those in the \condBase{} condition ($M = 16.8$), neither of which were significantly different from the trust rating of those in the \condHalf{} condition ($M = 20.1$). \autoref{fig:qualitya} shows this result in more detail. 

\noindent
\textbf{Explanation Satisfaction}: %Similar to model trust as discussed above, we summed seven Likert items (all with positive valance) adapted from Hoffman et al.~\cite{hoffman2018metrics} into a single rating of explanation satisfaction ranging from 7-49. 
Condition had a significant impact on satisfaction with the explanation ($F(2,80)=7.3$, $p=0.001$). A post-hoc test found that participants in the \condAll{} condition rated their satisfaction significantly higher ($M = 34.2$) than those in the other two conditions ($M = 26.4$) (see \autoref{fig:quality}).

\begin{figure*}
    \centering
    \includegraphics[width=0.8\textwidth]{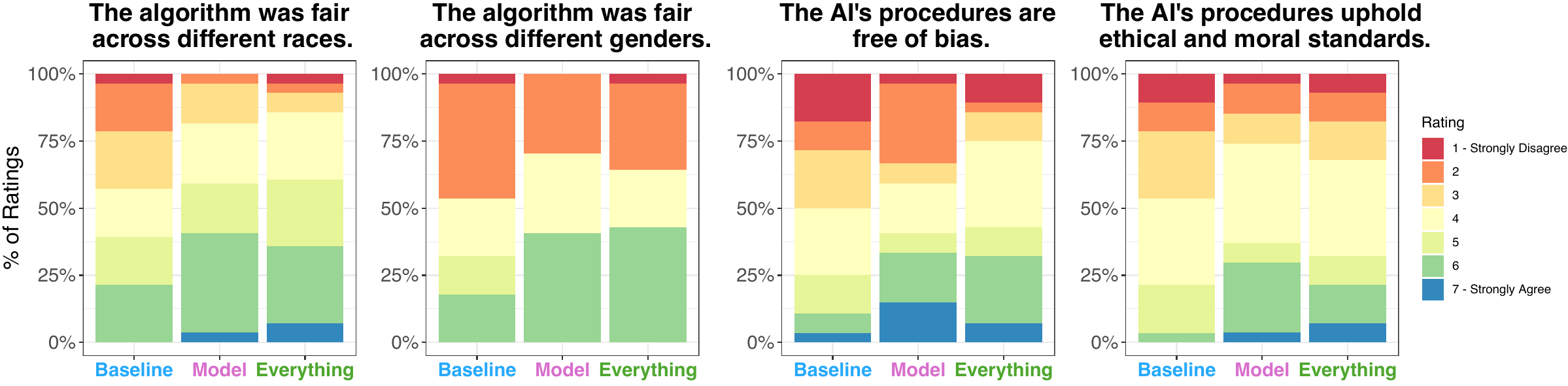}
    \caption{Likert responses from our participants when asked to assess potential biases in race and gender, or overall assessments of the ethics and biases of the model. While there was diversity in responses, in all cases on average, increasing the amount of information resulted in marginally higher ratings of perceived fairness (or lack of bias), despite the information being irrelevant to model fairness and bias.}
    \label{fig:likert}
\end{figure*}

\noindent
\textbf{Fairness}: %Individual Rating Items}: 
We asked participants to rate their agreement (from 1: Strongly disagree to 7: Strongly agree) with four additional questions around fairness (see \autoref{fig:likert}), which were modeled after Goyal et al.~\cite{goyal_impact_2024}. For questions around perceived gender biases, overall bias, and overall ethics, we found no significant difference among conditions (for gender: $F(2,80)=0.74$, $p=0.48$; for bias in general: $F(2,80)=1.9$, $p=0.16$; and for ethics in general: $F(2,80)=2.1$, $p=0.13$). For perceived fairness with respect to race, we did find a significant effect of condition ($F(2,80)=3.5$, $p=0.035$): those in the \condBase{} condition rated the algorithm as the least fair across race ($M = 3.9$), followed by those in the \condAll{} ($M = 4.8$) condition and then the \condHalf{} ($M = 4.8$) condition.

The most troubling result regarding participants' perceptions of model fairness is that the percentage of participants who rated our unfair model as fair---either in general or specifically with respect to race and gender---was highest in the two conditions with the most explanatory information. Equally concerning is that these two conditions had the lowest percentage of participants who correctly identified the unfairness of the model (see \autoref{fig:likert}).

\begin{figure}[h!]
    \centering
    \includegraphics[width=0.8\columnwidth]{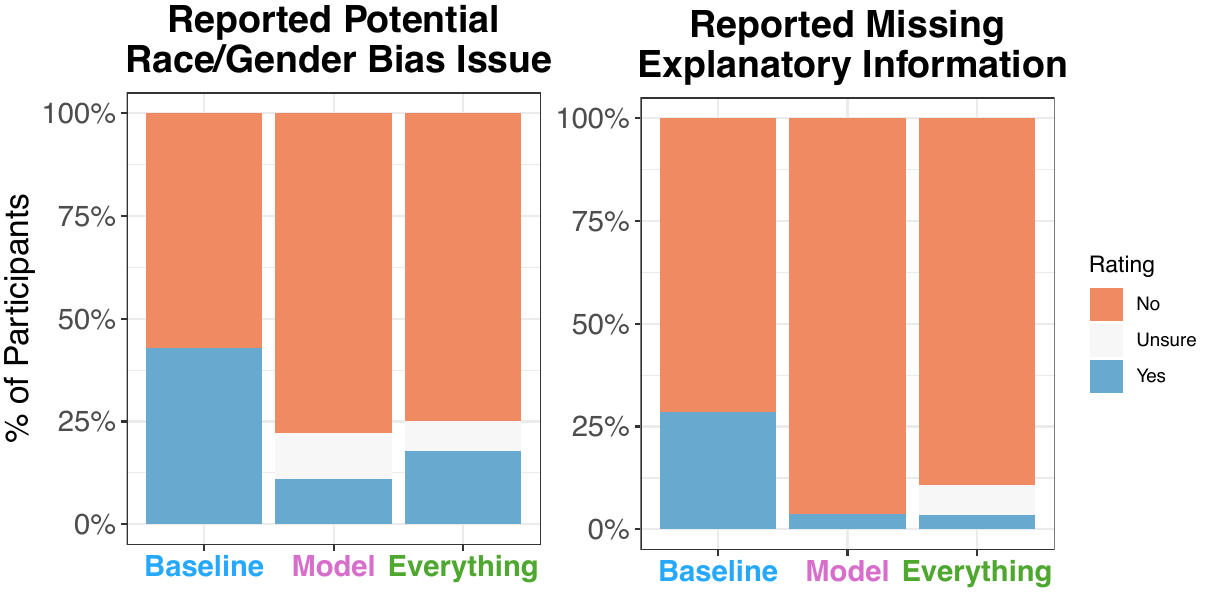}
    \caption{Frequency of our derived qualitative codes based on participants' free-text responses. Specifically, whether participants reported missing any explanatory information such as feature importance scores, fairness metrics, or model architecture (on the left); and whether participants reported any possible unfairness with respect to the race or gender (on the right). As a reminder, our model \textit{only} used race and gender to make decisions. The addition of uninformative explanatory components across our conditions resulted in lower rates of participants thinking more information was needed and noticing the severe biases in model outcomes.}
    \label{fig:qual}
\end{figure}
% \subsubsection{Qualitative Analyses}

\noindent
\textbf{Qualitative Findings}: Participants in the \condBase{} condition reported noticing \textit{more} issues with race or gender in the model's predictions ($12/28 = 42.9$\%) and gave \textit{more} responses indicating that important information was missing ($8/28 = 28.6$\%) compared to their counterparts in the \condHalf{} and \condAll{} conditions (see \autoref{fig:qual}).
% We closely examined free text responses to a question that was particularly relevant to understanding participants' mental models of our model's fairness: \textit{``What information about the candidate would you say was most important for the system in making its prediction?''}
% If a participant had an accurate mental model of the model's obvious unfair classifications, they would likely indicate in their open-ended response that the model relied primarily (or exclusively) on race and gender in its decision-making.
% We coded each response to identify whether the participant understood or alluded to the model's bias toward certain genders and races.
% We then counted how many participants demonstrated accurate mental models.
% The correct answer to this question was that the model exclusively used race and gender to make a decision, although we considered this answer unlikely given that our scenario described the model as using all available features. 
% Rather than the specific model internals, we were interested in whether participants detected any potential concerns related to the use of sensitive features.
% We counted a response as correct if both race and gender were mentioned (even if amongst other features), and partially correct if either race or gender were mentioned. 
Still, responses from only three participants (one from each condition) indicated a partial understanding of the information most important in guiding the model's predictions (e.g., P3 and P36 noted \textit{``race''} and \textit{``family income.''}). No response was fully correct. 
% For example, P2 said \textit{``I think the system looked at all of the candidate information as a whole and compared it to other passing candidates. If they matched other passing candidates (for instance, if they were male, while [White], with a lower GPA, that was in a bottom tier) the system could still say they would pass because other similar candidates passed,''} 
% P36 said \textit{``Race and family income,''} and P3 said \textit{``It appeared that family income and race were the main factors in determining if the student would pass the exam.''} We contrast the rarity of these statements with the more common responses from participants that downplay or dismiss the impact of race and gender on the model.  For example 

% Across all free text question responses, the second coder of this paper 
We noted 19 instances where participants explicitly described our model as fair (e.g., P8 said, ``\textit{It's fair in that it has no racial or gender bias.''}).  %P66 said, \textit{``The better the student's past academic success the better, such as GPA. The least important aspect was age, race and gender.''}).  %
% Despite participants' lack of success identifying our model's unfair reliance on race and gender, 
Still, participants occasionally expressed unease with the model. %, even 
P76 said they were concerned with \textit{``what kind of darkness people might use it for.''} %  Two of our participants, who self-identified as current or former law students, were particularly harsh. 
P52 stated:
\begin{quote}
\textit{``As a current law student I do not think this model is very fair at all. While it is true that things like the LSAT, GPAs, law school ranking, class ranking, etc. can be used to predict the likeliness of someone passing the bar, they are not perfect. I would say law school ranking and class ranking probably provide the best indicators since better schools have better bar passage rates as a fact, and better students tend to understand the subjects tested on the bar better, but someone can be at low tier school, ranked near the bottom of their class and still pass. A person is not just their data.''}
\end{quote}
We noted 15 instances of unease about the model's ability to holistically understand candidates.
% These quotes exemplify an emergent code, which the second coder reported occurring 15 times across all free text responses, where participants communicated unease about whether the model was capturing a holistic picture of the candidate, whether the model had sufficient data about individual perseverance, or how random chance could contribute to a candidate passing or failing the bar exam. 

\section{Discussion}
Our user study validates our central premise, that increasing amounts of nominally explanatory data can engender unwarranted trust in a predictive model.
We presented participants with explanations of an unfair and superficial model using \textit{only} race and gender to make a decision about academic success, even when much more informative features (such as GPA or class rank) were available. 
Regardless of their assigned condition, participants agreed with the model's predictions the majority of the time and often rated the model's quality highly.
This result persists even for our \condBase{} condition where the participants had, essentially, no information about the model other than eight predictions and the fact that it was made by computer scientists.

Consistent patterns of per-condition differences in participants' perceptions of our model, despite limited information, suggest the potential impact of ``trust junk'' is large. % to revisit our car analogy, if people are primed to buy the car before they step on the lot, then the salesman does not have to do very much persuasive work to make a sale.
%More relevant to the concept of bullshitting, though, were 
Our \condAll{} condition's explanation mostly reiterated information about the model's training data. It and the \condHalf{} condition's explanation include largely contextless global accuracy information.
Concerningly, the presence of this information, none of which provides insight on model internals or potential biases, resulted in increased rates of participant agreement with and trust in the model.
Only a minority of participants described the model as unfair or biased.
The few participants who did report concerns with the model were unable to accurately articulate \textit{why} the model was flawed, and, in the absence of crucial model information, resorted to informal, often incorrect reasoning to explain why the model might be wrong or unfair.

We additionally find that the \textit{amount} of information influenced fairwashing effects. Participants who saw all of the explanatory components (in the \condAll{} condition) demonstrated greater agreement with and reported higher trust in the model compared to those who saw the fewest explanatory components (in the \condBase{} condition).  They also reported fewer fairness concerns in their open-ended responses compared to their \condBase{} counterparts. In short, we found that trust junk \textit{worked}: participants were either ``soothed'' by the irrelevant information~\cite{weller2017challenges} or ``numbed'' by the sheer amount of data~\cite{correll_towards_2021} in the explanations.
All of this manipulation and persuasion occurred in the context of a model that was superficial and deeply unfair, precisely the case where we'd hope XAI would empower lay audiences to make accurate judgments.

Our tested conditions cover some, but not all, of the strategies that Wall et al.~\cite{wall2024trust} claim foster trust in viewers of XAI visualizations. Future work is needed to look at specifically how factors like disclosure of uncertainty information, aesthetic appeal, and other specific trust junk ``knobs'' interplay. What makes an XAI visualization persuasive, misleading, or overwhelming is likely to be a complex combination of many sociotechnical factors~\cite{ehsan2023charting} not amenable to the sort of self-contained study presented in this work.

While some of the issues we uncover can be addressed by increased data literacy in audiences, education alone is not enough (even self-described AI experts habitually misinterpret or fail to understand XAI techniques~\cite{kaur2020interpreting}).
We echo the assertion of Hullman et al.~\cite{hullman2025explanations}: the success of AI explanations can only be fairly assessed when considering the goals of such explanations. We urge the community to attend to the \textit{rhetorical} goals of XAI explanations and their manipulative power as intrinsically persuasive artifacts.

%% \section{Introduction} %for journal use above \firstsection{..} instead

\acknowledgments{
We thank Lace Padilla for comments on a draft of this work.}

\bibliographystyle{abbrv-doi}

\bibliography{template}
\end{document}